\documentclass{article}
\usepackage{amsmath,epsfig}
\usepackage[preprint]{spconfa4}
\usepackage{subfig}
\usepackage{booktabs}
\usepackage{marvosym}
\usepackage{multirow}
\usepackage{makecell}
\usepackage{color}

\usepackage{fancyhdr}

\let\OLDthebibliography\thebibliography
\renewcommand\thebibliography[1]{
  \OLDthebibliography{#1}
  \setlength{\parskip}{0pt}
  \setlength{\itemsep}{0pt plus 0.3ex}
}

\begin{document}\sloppy

\def\x{{\mathbf x}}
\def\L{{\cal L}}

\title{Subjective Quality Assessment for Images Generated by \\ Computer Graphics}
%
\name{Tao Wang$^*$,Zicheng Zhang$^*$,Wei Sun,Xiongkuo Min,Wei Lu,and Guangtao Zhai}
\address{Institute of Image Communication and Network Engineering, Shanghai Jiao Tong University, China \\
zzc1998@sjtu.edu.cn}

 \maketitle

\let\thefootnote\relax\footnotetext{ $^*$These authors make equal contributions to this work.}

\begin{abstract}
With the development of rendering techniques, computer graphics generated images (CGIs) have been widely used in practical application scenarios such as architecture design, video games, simulators, movies, etc. Different from natural scene images (NSIs), the distortions of CGIs are usually caused by poor rending settings and limited computation resources. What's more, some CGIs may also suffer from compression distortions in transmission systems like cloud gaming and stream media. However, limited work has been put forward to tackle the problem of computer graphics generated images' quality assessment (CG-IQA). Therefore, in this paper, we establish a large-scale subjective CG-IQA database to deal with the challenge of CG-IQA tasks. We collect 25,454 in-the-wild CGIs through previous databases and personal collection. After data cleaning, we carefully select 1,200 CGIs to conduct the subjective experiment. Several popular no-reference image quality assessment (NR-IQA) methods are tested on our database. The experimental results show that the handcrafted-based methods achieve low correlation with subjective judgment and deep learning based methods obtain relatively better performance, which demonstrates that the current NR-IQA models are not suitable for CG-IQA tasks and more effective models are urgently needed. 
\end{abstract}
%
\section{Introduction}
Unlike natural scene images (NSIs), which are captured from the real world, computer graphic images (CGIs) are artificially rendered using computer graphics \cite{min2017unified}. 
More specifically, rendering indicates the process of generating images from a 2D or 3D model described by means of a computer program, which has been widely used in architecture, video games, simulators, movies, etc. \cite{pharr2016physically}.
Note that we do not consider the images generated by the neural network based generative models such as GAN \cite{goodfellow2014generative} and NERF \cite{mildenhall2020nerf} since the quality of these images are largely dependent on training samples and well-designed networks, which is out of our research purposes. Details of the quality assessment for generative images can be obtained from \cite{borji2019pros,gu2020giqa,jinjin2020pipal}.
According to the rendering tools and purposes, CGIs can be further categorized into photorealistic images and non-photorealistic images.
Photorealistic images are rendered by modeling the real world to achieve photorealism \cite{greenberg1997framework}, while non-photorealistic images focus on enabling a wide variety of expressive styles for digital art, such as painting, drawing, and animated cartoons  \cite{card2002non}. Despite of the different rendering purposes and techniques, both photorealistic and non-photorealistic images share similar distortions, such as texture loss caused by the limited computation resources, poor visibility caused by wrong exposure setting, and blur caused by low rendering accuracy. What's more, accompanied by the rapid development of network service and entertainment consumption, people can perceive large numbers of CGIs through live broadcast and cloud gaming \cite{laghari2019quality}. CGIs in such situations are also affected by the inevitable compression distortions in transmission systems, which damage the quality of CGIs and greatly influence user's Quality of Experience (QoE).

\begin{table*}[]
\centering
\caption{The comparison of previous CGI related databases and our database}
\vspace{-0.2cm}
\begin{tabular}{ccccc}
\toprule
Database       & Scale & Source                & Scope             & Resolution Range     \\
\midrule
CCT \cite{min2017unified}           & 528   & PC game images        & Compression distortion       & 720P 1080P           \\
TGQA \cite{ling2020subjective}          & 1091  & Mobile game images    & Aesthetic evaluation         & 1080P                  \\
KUGVD \cite{barman2019no} & 90    & PC game videos      & Downsampling \& Bitrates control  &480P 720P 1080P       \\
TGV \cite{wen2021subjective}           & 1293  & Mobile game videos    & Stull \& Bitrates control              & 480P 720P 1080P  \\
Ours           & 1200  & 3D movie \& 3D game images & Rendering effect \& Compression distortion      & 480P$\sim$4K   \\
\bottomrule
\end{tabular}
\vspace{-0.4cm}
\label{tab:compare}
\end{table*}


     In the last decade, image quality assessment (IQA) has been developed to facilitate the progress of image processing and analysis. Large numbers of IQA models \cite{min2017unified,mittal2012no,min2018blind,saad2012blind,gu2014using,zhang2018blind,su2020blindly,ke2021musiq,zhang2021no} have been proposed to predict the human perception of NSIs. However, CGIs and NSIs have huge differences in content and distortions, which indicates that it is difficult to transfer previous NSI-specific models to the CGI field. What's more, little work has been dedicated to specifically assessing the quality of CGIs. A summarization of the gap in CGIs quality assessment (CG-IQA) is given here: 1) Nearly all existing CG-IQA databases are constructed in former times. The CGIs are selected from limited types of sources. Such CGIs are outdated and not able to cover the range of current rendering techniques. 2) The existing CG-IQA databases are relatively small in scale, which are not sufficient to support deep learning methods. 3) Most of the developed methods are full-reference (FR) and focus on limited types of distortions. Such methods focus on the quality of CGIs whose distortions are manually introduced to the reference images. However, in most situations, the reference CGIs are not available and the types of distortions are various, meaning the previous methods may not meet the needs of practical applications.


Therefore, to overcome the shortcomings of the existing databases, we propose a new database to tackle the challenge of CG-IQA. The proposed database collects 1,200 in-the-wild images from two typical scenes (3D movie and 3D game) and we conduct the subjective experiment to obtain mean opinion scores (MOS) of CG images. Table. \ref{tab:compare} illustrates the difference between our database and other databases with the similar scope. Our database has the following advantages:  1) Our database is sourced from both 3D movies and 3D games, thus obtaining high diversity. 2) We assess the quality of CGIs from both aspects of rendering artifacts and compression distortions. Additionally, such compression distortions are caused by real-world applications and we do not manually introduce new compression distortions. 3) Our database covers a wider range of resolutions, which helps our database gain more generality and ensures our work meets most screen resolution needs in current times.  Besides, we employ several successful no-reference image quality assessment (NR-IQA) metrics to compare the performance on the database.


\section{Subjective Quality Assessment}
\subsection{CGIs Collection}
Thanks to the contributions of the LSCGB database \cite{bai2021robust} (a database designed for computer-generated images forensic), we collect 9,212 3D game images and 8,818 3D movie images sourced from 18 popular 3D games and 26 recent 3D movies. Additionally, we personally collect 3,424 3D games images from screenshots of local game demos and cloud gaming to increase content diversity and cover more range of resolutions. We also personally collect 4,000 3D movie images by extracting the frames of the movies. To ensure the scene diversity and restrict the distractions of irrelevant things, we specifically eliminate the CGIs with similar contents and remove the life bars, mini maps, and subtitles. Considering that most CGIs are obtained through compressed streams or videos except screenshots from local game demos, we do not manually introduce new compression distortions. After the process described above, a total of 1,200 CGIs are obtained, in which 600 CGIs are from 3D games and 600 CGIs are from 3D movies.  

\begin{figure*}[]
\centering
\subfloat[MOS=11.06]{\includegraphics[width=0.2\textwidth]{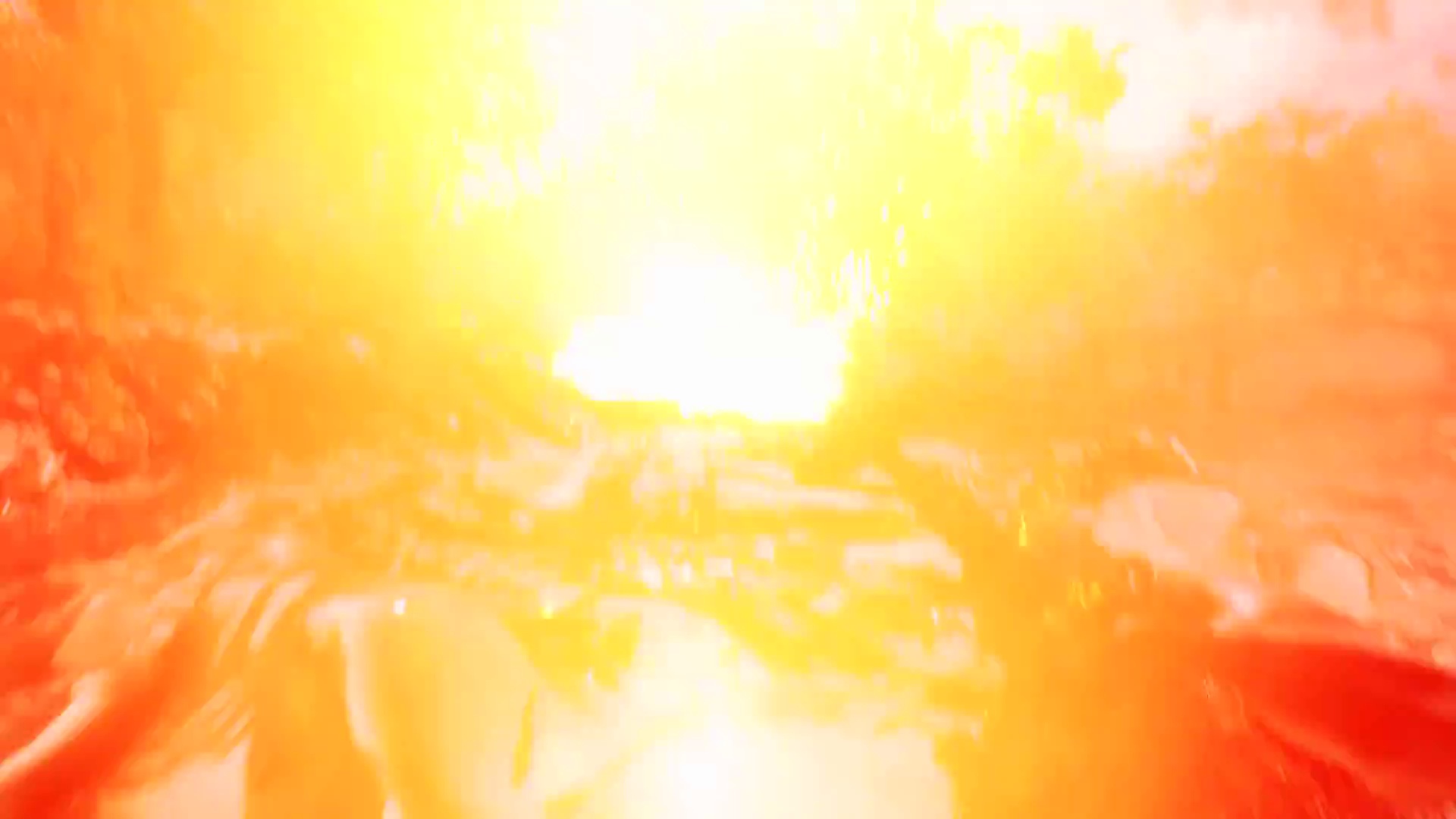}%
}
\subfloat[MOS=42.43]{\includegraphics[width=0.2\textwidth]{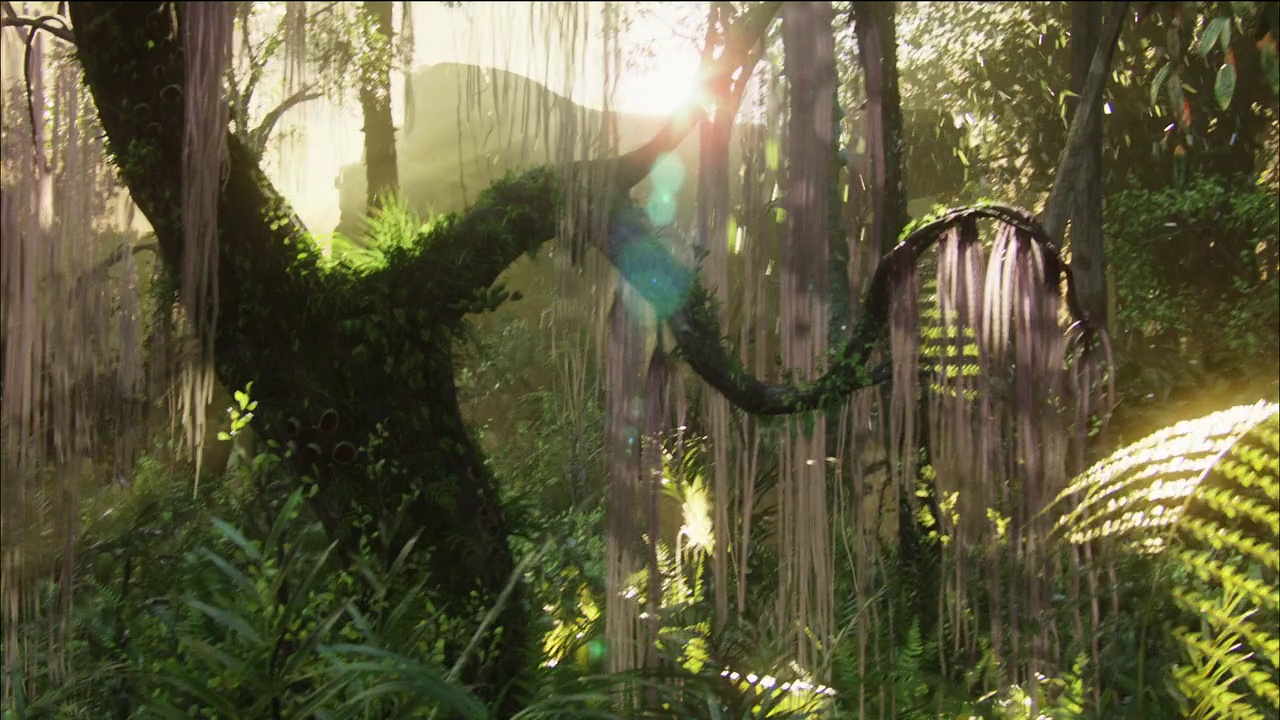}%
}
\subfloat[MOS=51.70]{\includegraphics[width=0.2\textwidth]{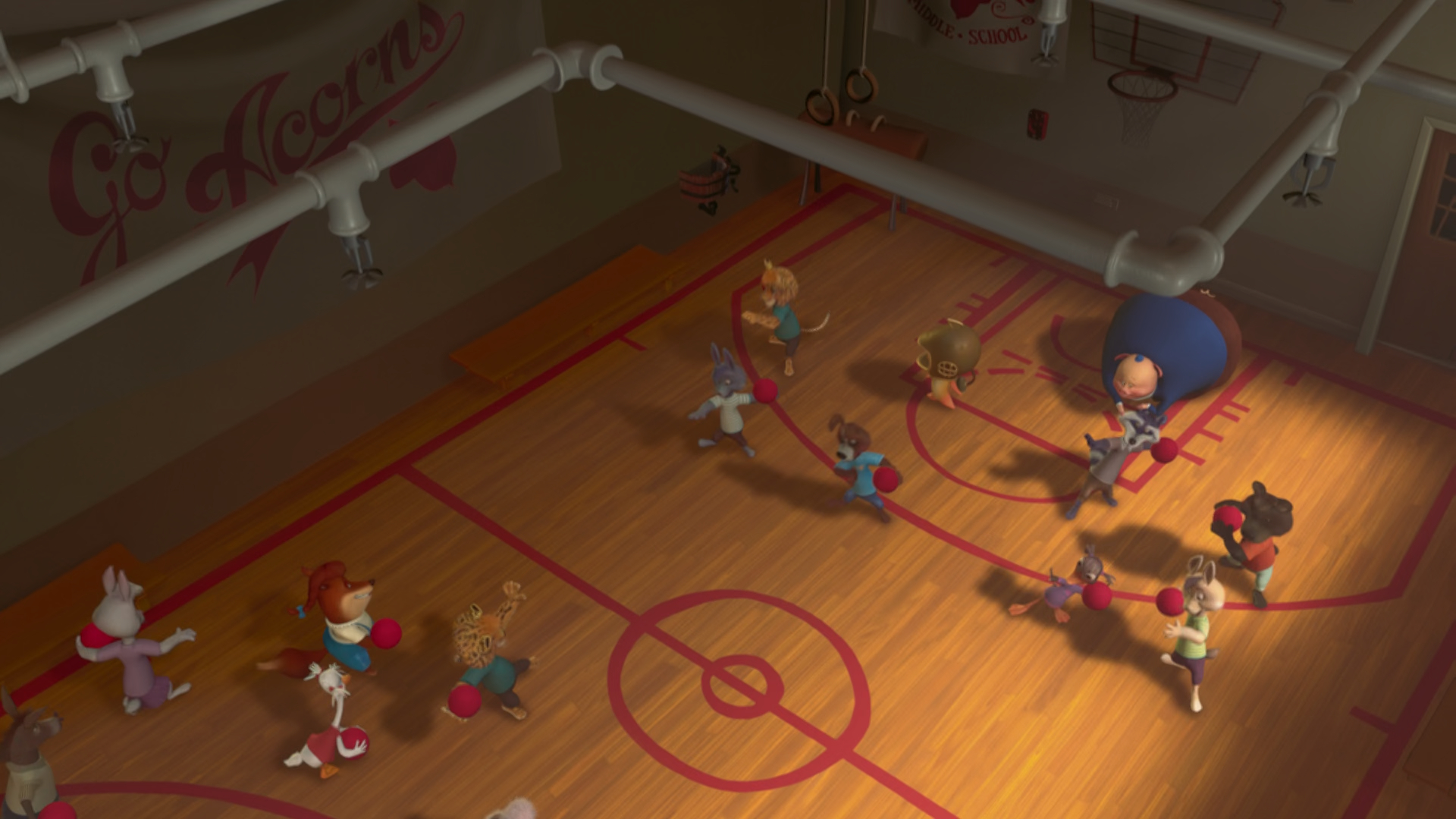}%
}
\subfloat[MOS=62.49]{\includegraphics[width=0.2\textwidth]{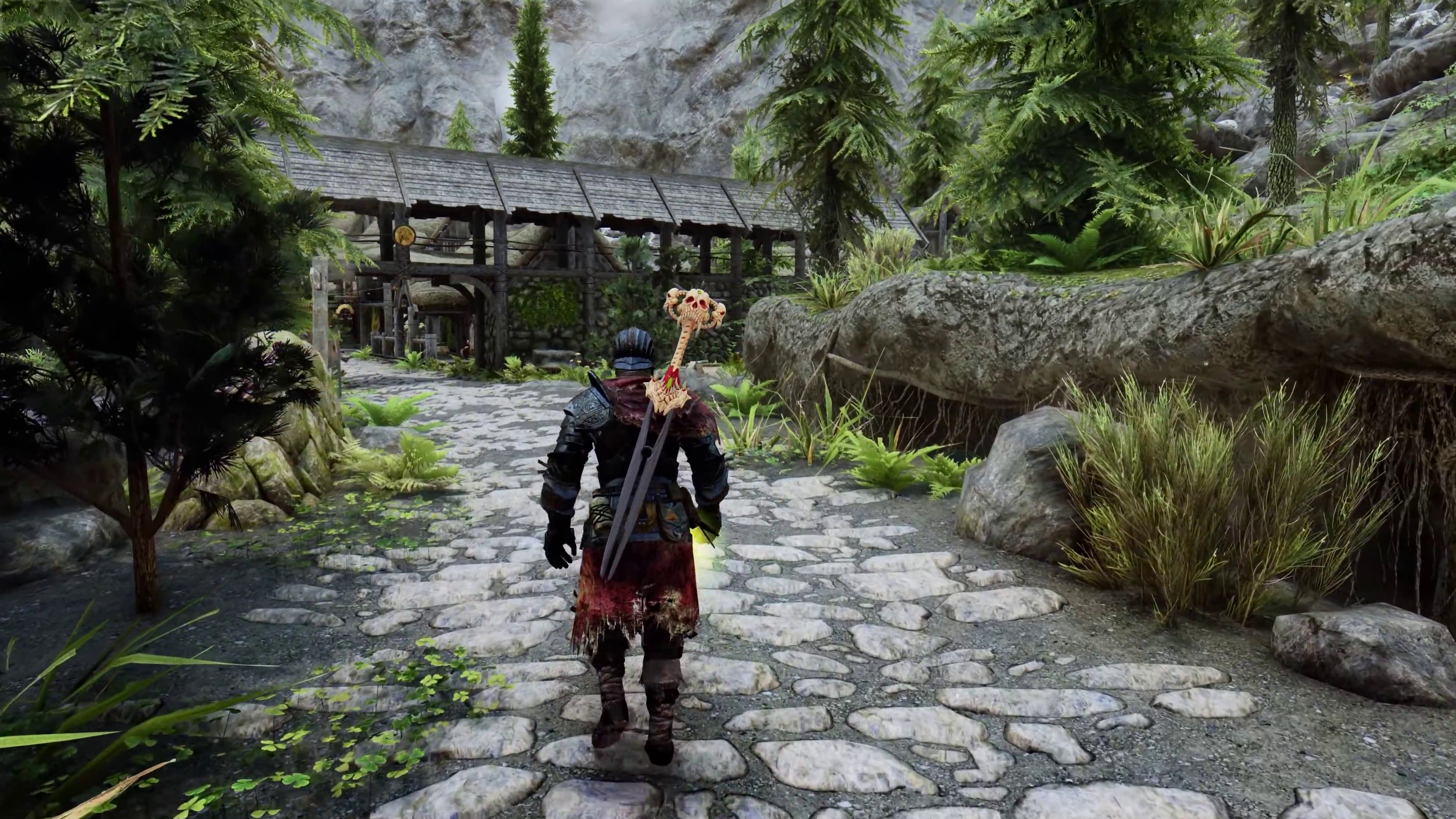}%
}
\subfloat[MOS=76.85]{\includegraphics[width=0.2\textwidth,height=0.1125\textwidth]{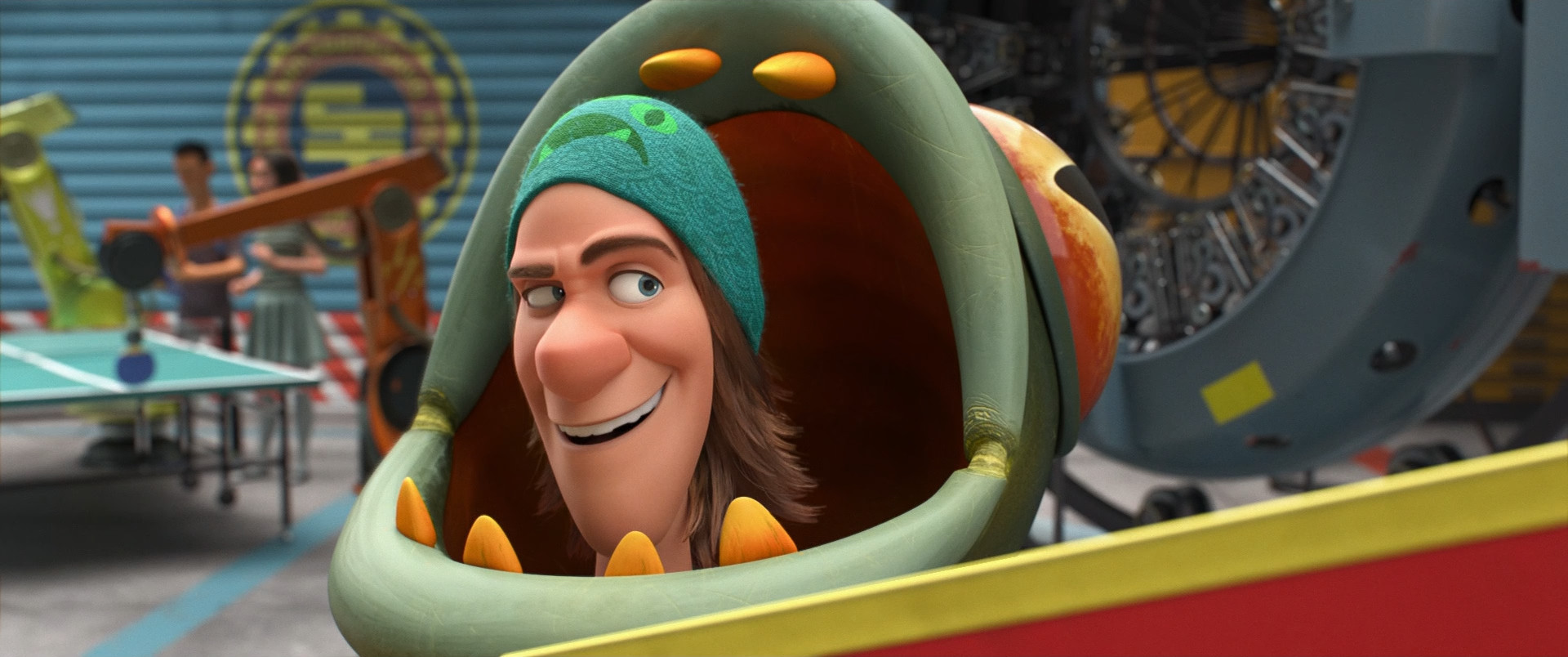}%
}
\vspace{-0.2cm}
\caption{Examples from our database and their corresponding MOS.}
\vspace{-0.4cm}
\label{fig:example}
\end{figure*}

\subsection{Subjective Experiment Methodology}
To get the subjective quality values of CGIs, we conduct the subjective experiment based on the recommendations of ITU-R BT.500-13 \cite{bt2002methodology}. All the CGIs are shown in random order with an interface designed by Python Tkinter on an iMac monitor which supports the resolution up to 4096 $\times$ 2304. 
We invite 12 male subjects and 8 female subjects to participate in the subjective experiment. The viewers are seated at a distance of around 1.5 times the screen height (45cm) in a laboratory
environment that has normal indoor illumination levels. The quality scale scores from 0 to 5, with a minimum interval of 0.1. The whole experiment is split into four sessions and each session includes 300 CGIs for quality evaluation to ensure that each session lasts no more than half an hour. Therefore, each CGI is evaluated by 20 subjects, which generates a total of 20$\times$1,200=24,000 quality scores. 

After the subjective experiment, we obtain all the quality scores from the subjects. Let $r_{ij}$ denote the raw score provided by the $i$-th subject on the $j$-th image, the z-scores are computed from the raw scores as follows:
\begin{equation}
z_{i j}=\frac{r_{i j}-\mu_{i}}{\sigma_{i}},
\end{equation}
where $\mu_{i}=\frac{1}{N_{i}} \sum_{j=1}^{N_{i}} r_{i j}$, $\sigma_{i}=\sqrt{\frac{1}{N_{i}-1} \sum_{j=1}^{N_{i}}\left(r_{i j}-\mu_{i}\right)}$, and $N_i$ is the number of images seen by subject $i$.
After that, we remove scores from unreliable subjects by using the recommended subject rejection procedure in the ITU-R BT.500-13 \cite{bt2002methodology}.
The z-scores are then linearly rescaled to $[0,100]$. Finally, the MOS of the image $j$ is calculated by averaging the rescaled z-scores: 
\begin{equation}
M O S_{j}=\frac{1}{M} \sum_{i=1}^{M} z_{i j}^{'},
\end{equation}
where $M O S_{j}$ indicates the MOS for the $j$-th CGI, $M$ is the number of the valid subjects, and $z_{i j}^{'}$ are the rescaled z-scores. Some CGI samples and their corresponding MOSs are shown in Fig. \ref{fig:example}. It can be seen that invisibility caused by wrong exposure and blur tend to result in poor MOS while CGIs containing exquisite details and fine illumination settings are more preferred by the viewers.

\begin{figure}
    \centering
    \includegraphics[width=\linewidth]{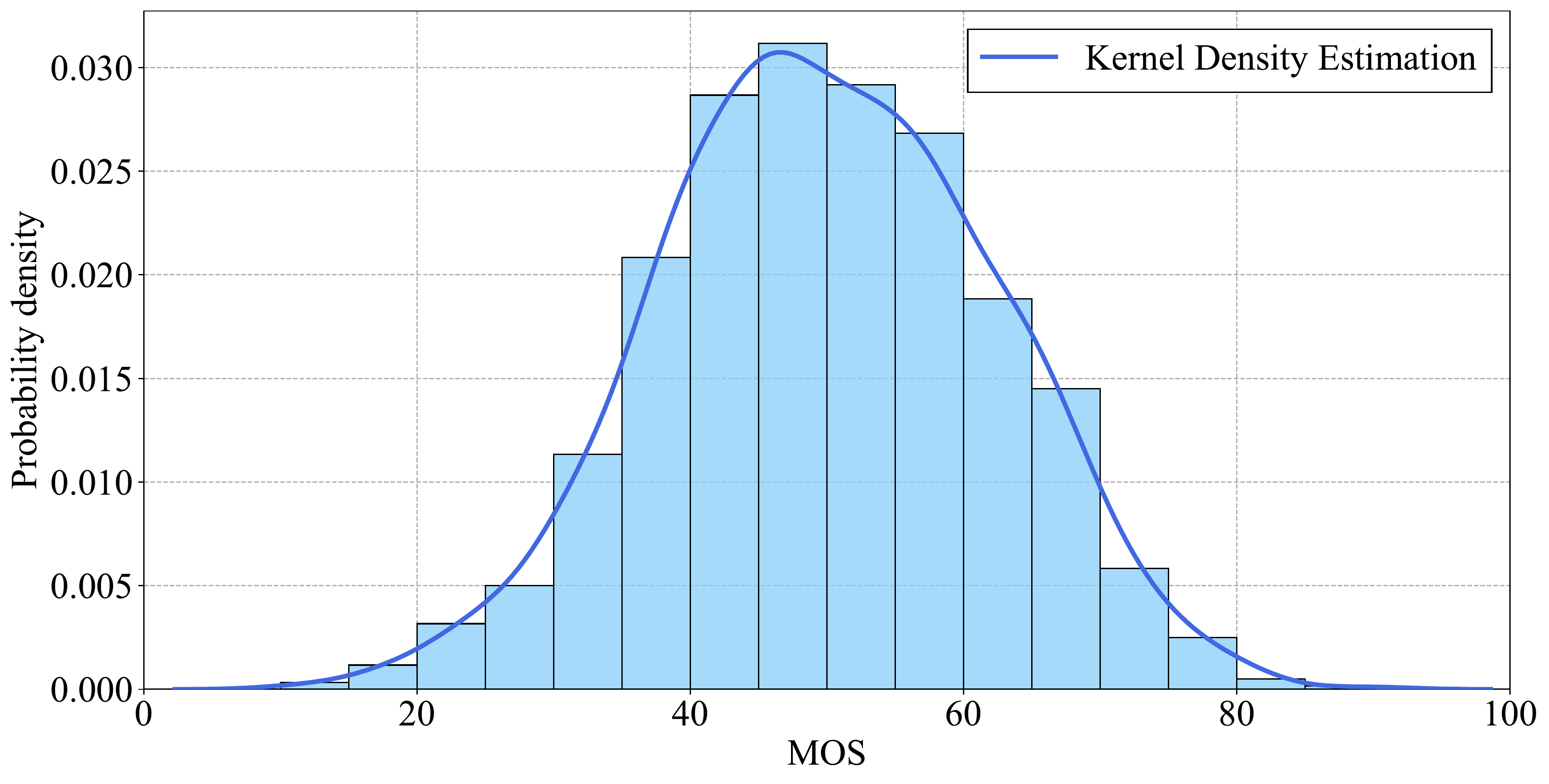}
    \vspace{-0.2cm}
    \caption{The distribution of the MOSs of the constructed database.}
    \label{fig:mos}
    \vspace{-0.4cm}
\end{figure}

\begin{figure}
    \centering
    \includegraphics[width=\linewidth]{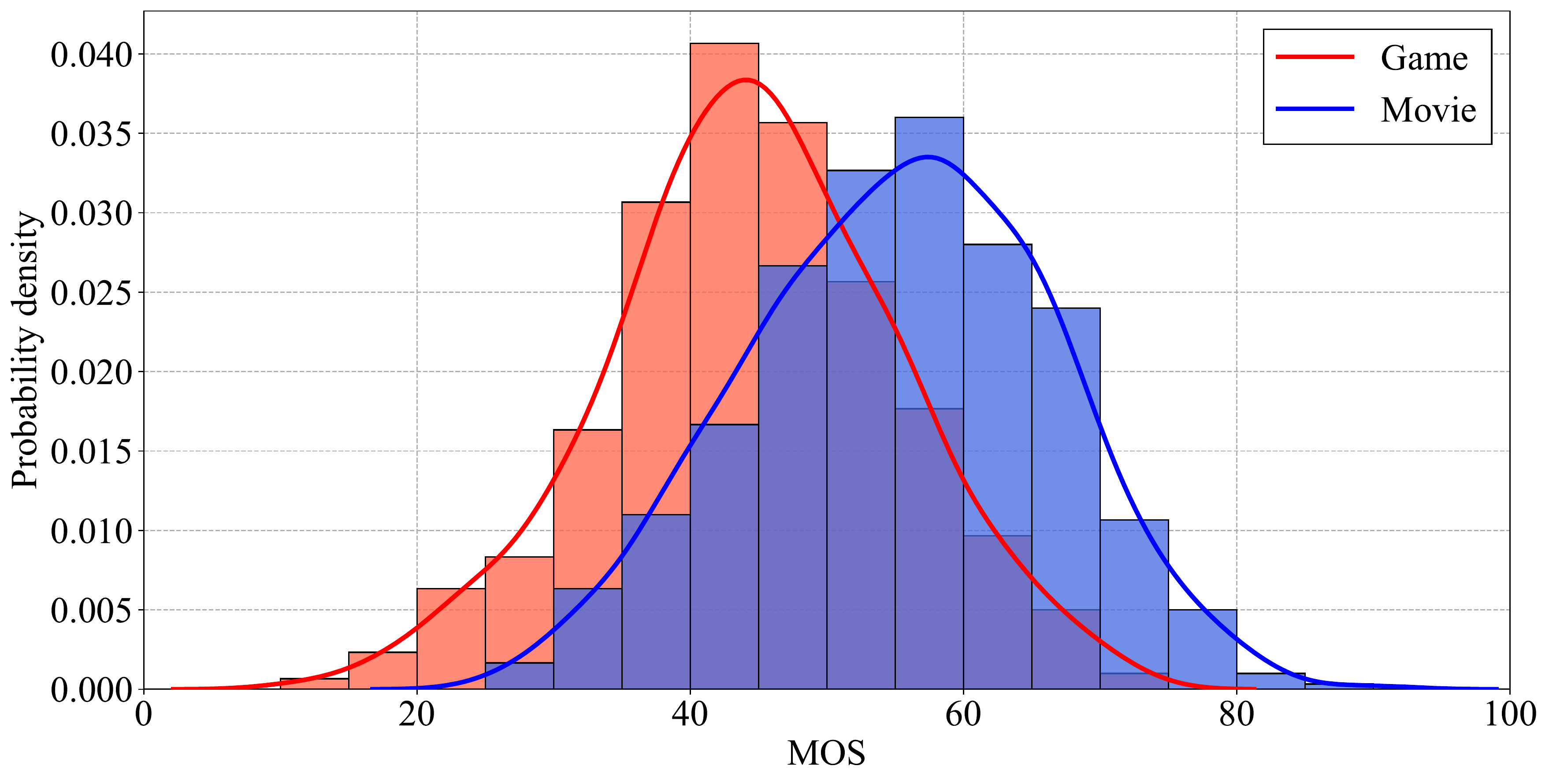}
    \vspace{-0.2cm}
    \caption{The MOS distribution of 3D  movies images and 3D games images separately.}
    \label{fig:smos}
    \vspace{-0.4cm}
\end{figure}

The distribution of the obtained MOS is illustrated in Fig. \ref{fig:mos} and the respective distributions of MOS for movies images and games images are exhibited in Fig. \ref{fig:smos}, from which we can make several observations. First, the MOS of most images are located in the range [20, 80], showing a Gaussian-like distribution. Second, the score of images from 3D movies is slightly higher than that of images from 3D games in general, which is in line with common sense that movie content is carefully rendered through many complex processes, while game content is generated in real-time with limited sources.



\section{Performance and Analysis}
Many IQA methods have been proposed to accurately predict the quality levels of distorted images in the last decade. Considering that the CGIs in our database do not have pristine sources, only no-reference image quality assessment (NR-IQA) models are qualified for evaluating their quality. Hence we select several representative NR-IQA models to test their effectiveness on our database.



\begin{table*}[]
\centering
  \caption{Performance comparison of state-of-the-art NR-IQA methods on the proposed database. 
The best-performing method in each row is highlighted.}
\vspace{-0.2cm}
  \label{tab:all}
\resizebox{\textwidth}{!}{
\begin{tabular}{c|ccccc|ccccc}

\toprule
Type & \multicolumn{5}{c|}{Handcrafted-based} & \multicolumn{5}{c}{Deep learning based}     \\
\hline
 Methods & BRISQUE & BMPRI &BLIINDS  & NFERM & UCA  & DBCNN &HyperIQA  & MUSIQ  & MGQA & StairIQA \\
\hline
SRCC $\uparrow$    & 0.2669 & 0.1870  &  0.1644  &  0.1502 &  0.1824  &    0.5868 &  0.6966 & 0.6907 & 0.7013 & \textbf{0.7199}   \\
PLCC $\uparrow$   & 0.2640 & 0.1926  &  0.1700  &  0.1330  &  0.2021  &    0.5893 &   0.6989 & 0.6883 & 0.7194 & \textbf{0.7276}  \\
KRCC $\uparrow$   & 0.1804  & 0.1255 &  0.1117 &   0.1010  &  0.1224 &     0.4186 &   0.5103   & 0.5010  & 0.5154 & \textbf{0.5300}  \\
RMSE $\downarrow$   & 11.8700 & 11.9551 &  11.8996  &  12.0222  &  11.9603  &  10.8278 &  8.8676    & 10.0001 & 8.9580 & \textbf{8.5841}  \\
\bottomrule
\end{tabular}}
\vspace{-0.4cm}
\end{table*}

\subsection{Comparing Algorithms and Evaluation Criteria}

On the proposed database, we list the performance of multiple state-of-the-art (SOTA) NR-IQA algorithms, which can be categorized into two types: \\
$\bullet$ Handcrafted-based methods: BRISQUE \cite{mittal2012no}, BMPRI \cite{min2018blind}, BLII\-NDS-II \cite{saad2012blind}, NFERM \cite{gu2014using} and UCA \cite{min2017unified}. These methods operate by extracting handcrafted features from images and regress the features to quality scores. Besides, UCA is an opinion-unware method and do not need training. \\
 $\bullet$ Deep learning based methods:  DBCNN \cite{zhang2018blind}, HyperIQA \cite{su2020blindly}, MUSIQ \cite{ke2021musiq}, MGQA \cite{wang2021multi} and StairIQA \cite{sun2021blind}. Additionally, MUSIQ is a transformer-based model.



To quantitatively show the assessment ability of different methods, four mainstream evaluation criteria are utilized to compare the performance between the predicted scores and MOS, which include Spearman Rank Correlation Coefficient (SRCC), Kendall’s Rank Correlation Coefficient (KRCC), Pearson Linear Correlation Coefficient (PLCC), Root Mean Squared Error (RMSE). An excellent model should obtain values of SRCC, KRCC, and PLCC close to 1, and the value of RMSE near 0. Before calculating the criteria, we first apply a five-parameter logistic function to map the predicted scores according to the practices in \cite{sheikh2006statistical}:
\begin{equation}
\hat{y}=\beta_{1}\left(0.5-\frac{1}{1+e^{\beta_{2}\left(y-\beta_{3}\right)}}\right)+\beta_{4} y+\beta_{5}
\end{equation}
where $\left\{\beta_{i} \mid i=1,2, \ldots, 5\right\}$ are parameters to be fitted, $y$ and $\hat{y}$ are the predicted scores and mapped scores respectively. 

\subsection{Experiment Setup}
Normally speaking, most NR-IQA methods are training based, which needs a training set to learn a mapping function between the feature/image domain and the quality domain. Thus we spilt the database with a ratio of 8:2 for the training set and testing set respectively. 
For deep learning based methods, we \textbf{retrained} all the models on the new database with default hyperparameter settings for validation.

To reduce the effect of randomness, we repeat the above process 1,00 times for handcraft-based methods and 10 times for deep 
learning based methods, and record the average values of SRCC, KRCC, PLCC, and RMSE as the final experimental results.

\subsection{Experiment Performance}
\subsubsection{Handcrafted-based Methods}
The experimental results are clearly shown in Table \ref{tab:all}. With closer inspection, we can find that all the mainstream handcrafted-based NR-IQA methods designed for NSIs are not effective for predicting the quality value of CGIs. BRISQUE \cite{mittal2012no} is a classic NR-IQA model based on natural scene statistics (NSS) and obtains the best performance among handcrafted-based methods. However, its SRCC and PLCC results are still lower than 0.3. We attempt to give the reasons for the poor performance of the above handcrafted-based NR-IQA methods.

1) The mentioned handcrafted-based methods are specially designed for NSIs and NSIs are quite different from CGIs in content. The majority of them usually utilize many NSS distributions to estimate quality-aware parameters and these parameters do not work for CGIs, since the attribute distributions of CGIs do not meet the prior knowledge of NSIs. In all, the knowledge learned from NSIs by such NR-IQA methods is not suitable for qualifying the quality of CGIs. 

2) The CGIs in our database are more diverse in distortion types and resolutions. The selected handcrafted-based methods are designed on traditional databases where only one or several distortions are manually added to the reference images and the range of resolutions is limited. Therefore, it is not surprising that these methods achieve poor performance on our database.

\subsubsection{Deep Learning Based Methods}
The selected deep learning based NR-IQA methods are significantly superior to handcrafted-based NR-IQA methods and StairIQA \cite{sun2021blind} achieves first place in all criteria. However, the SRCC and PLCC results of StairIQA are lower than 0.75, which are not satisfactory enough. We also try to analyze the reasons for deep learning based methods' performance.

1) Deep learning based methods have a better ability to extract features that meet the training scope, which enables them to gain better performance. However, the backbones employed in such models are all pre-trained on databases consisting of mostly NSIs (e.g. ImageNet), which may limit the effectiveness of deep learning based models.

2) We think the scale of our database is still not enough to reach the performance bottleneck of deep learning based methods and they may suffer from the effect of over-fitting. Therefore, we will push forward our work and enlarge the scale of our database for further research.

\section{conclusion}
Computer graphics generated images (CGIs) are becoming more and more common in people's entertainment life, which makes the quality assessment of CGIs a hot topic in the multimedia area. In this paper, we conduct a large-scale subjective study for the quality assessment of CGIs and create a new CG-IQA database. 1200 images are selected from dozens of different genres of 3D games and 3D movies for the construction of our database. In subjective experiments, a total of 24,000 subjective scoring data are collected from 20 observers. Several popular SOTA NR-IQA methods also test on the new database, which include both handcrafted-based and deep learning based methods. The experimental results show that the handcrafted-based methods achieve low correlation with human perception and deep learning based methods achieve relatively better performance, but are far from satisfactory.
Our proposed database fills a gap in the research field for CG-IQA. In future work, we would further enlarge the scale of the database and carry out specific IQA models for CG-IQA problems.
\bibliographystyle{IEEEbib}
\bibliography{icme2021template}

\end{document}